\documentclass[sigconf]{acmart}
\AtBeginDocument{%
  }

\setcopyright{acmlicensed}
\copyrightyear{2018}
\acmYear{2018}
\acmDOI{XXXXXXX.XXXXXXX}
\acmConference[Conference acronym 'XX]{Make sure to enter the correct
  conference title from your rights confirmation email}{June 03--05,
  2018}{Woodstock, NY}
\acmISBN{978-1-4503-XXXX-X/2018/06}

\usepackage{multirow}
\begin{document}

\title{DiffusionGS: Generative Search with Query Conditioned Diffusion in Kuaishou}

\author{Qinyao Li}
\authornote{Work done during an internship at Kuaishou Technology.}
\affiliation{%
  \institution{Kuaishou Technology}
  \city{Hangzhou}
  \country{China}}
\email{qinyaoli369@gmail.com}

\author{Xiaoyang Zheng}
\orcid{0000-0001-5090-9819}
\authornote{*corresponding author}
\affiliation{%
  \institution{Kuaishou Technology}
  \city{Hangzhou}
  \country{China}}
\email{zhengxiaoyang@kuaishou.com}

\author{Qihang Zhao}
\affiliation{%
  \institution{Kuaishou Technology}
  \city{Hangzhou}
  \country{China}}
\email{zhaoqh75@mail.ustc.edu.cn}

\author{Ke Xu}
\authornote{*corresponding author}
\affiliation{%
  \institution{City University of Hong Kong}
  \city{Hong Kong}
  \country{China}}
\email{kkangwing@gmail.com}

\author{Zhongbo Sun}
\affiliation{%
  \institution{Kuaishou Technology}
  \city{Hangzhou}
  \country{China}}
\email{sunzb17@gmail.com}

\author{Chao Wang}
\affiliation{%
  \institution{Kuaishou Technology}
  \city{Hangzhou}
  \country{China}}
\email{zhangcong07@kuaishou.com}

\author{Chenyi Lei}
\orcid{0000-0001-6287-3673}
\affiliation{
  \institution{Kuaishou Technology}
  \city{Beijing}
  \country{China}}
\email{leichy@mail.ustc.edu.cn}

\author{Han Li}
\affiliation{
  \institution{Kuaishou Technology}
  \city{Beijing}
  \country{China}}
\email{lihan08@kuaishou.com}

\author{Wenwu Ou}
\affiliation{
  \institution{Unaffiliated}
  \city{Beijing}
  \country{China}}
\email{ouwenwu@gmail.com}
\renewcommand{\shortauthors}{Trovato et al.}
\newcommand{\kk}[1]{\textcolor[rgb]{1,0,0}{#1}}
\begin{abstract}
Personalized search ranking systems are critical for driving engagement and revenue in modern e-commerce and short-video platforms.
While existing methods excel at estimating users’ broad interests based on the filtered historical behaviors, they typically under-exploit explicit alignment between a user’s real-time intent (represented by the user query) and their past actions.
In this paper, we propose {\bf DiffusionGS}, a novel and scalable approach powered by generative models. 
Our key insight is that user queries can serve as explicit intent anchors to facilitate the extraction of users’ immediate interests from long-term, noisy historical behaviors.
Specifically, we formulate interest extraction as a conditional denoising task, where the user’s query guides a conditional diffusion process to produce a robust, user intent-aware representation from their behavioral sequence.
We propose the User-aware Denoising Layer (UDL) to incorporate user-specific profiles into the optimization of attention distribution on the user’s past actions.
By reframing queries as intent priors and leveraging diffusion-based denoising, our method provides a powerful mechanism for capturing dynamic user interest shifts.
Extensive offline and online experiments demonstrate the superiority of {\bf DiffusionGS} over state-of-the-art methods.
\end{abstract}

\ccsdesc[500]{Information systems~Content ranking}

\keywords{Generative Search Model; Diffusion Model; Conditioned Generation}

\received{20 February 2007}
\received[revised]{12 March 2009}
\received[accepted]{5 June 2009}

\maketitle

\section{Introduction}
User personalized search ranking systems play a pivotal role in both E-Commerce (e.g., Amazon and Taobao~\cite{zhou2018deep}) and short-video platforms (e.g., TikTok and Kuaishou~\cite{guo2023query}), as they directly influence user engagement and platform revenue.
To enhance user engagement and foster positive behaviors (e.g., click-throughs, conversions, and prolonged interactions) on platforms, the ranking systems are expected to rate and present documents within the candidate pool in a properly ranked order, through a comprehensive consideration of both the immediate user intent (articulated via the user query) and historical behavior sequences.
Therefore, accurately modeling the relationship between user query inputs and their historical behavior sequences is the key to ranking systems.

The advanced study of ranking models can be divided into two categories: Deep Learning Ranking Models ({DLRM}) and Generative Ranking Models ({GRM}).
The DLRM-based methods~\cite{zhou2018deep,zhou2019deep,si2024twin} typically rely on a two-stage approach, i.e., a General Search Unit (GSU) and an Exact Search Unit (ESU), where they design different attention mechanisms to extract discriminative user interest representations, while balancing the limited latency constraints of online systems with the substantial computational cost imposed by massive lifelong historical data.
Recently, the GRM-based methods~\cite{hou2024large,zhai2024actions,deng2025onerec} follow the advancement of Large Language Model (LLM) techniques to integrate the generative paradigm into search and recommendation systems.
These GRM works~\cite{hou2024large,zhai2024actions,deng2025onerec} typically organize user historical behaviors as tokens like natural languages, and employ transformer architectures to perform the next-token prediction task, for context-aware generation of user interest representations.

\begin{figure}
    \centering
    \includegraphics[width=\linewidth]{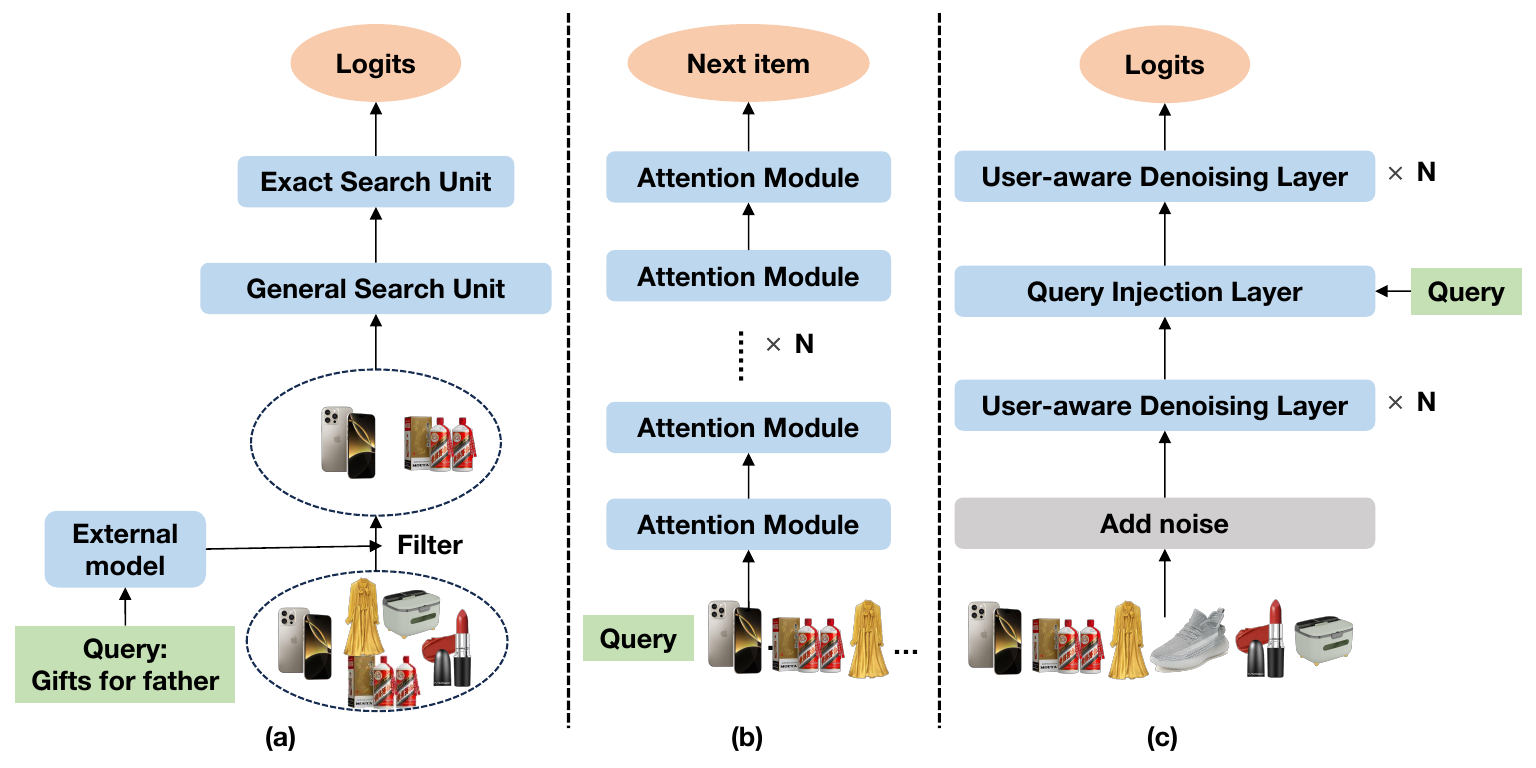}
    \caption{User history modeling in DLRMs, GRMs, and our DiffusionGS:
(a) DLRMs filter ``irrelevant'' items using queries, which may fail to capture underlying user preference patterns.
(b) GRMs concatenate queries as tokens with behavior sequences equally, where user intent may often be buried.
(c) DiffusionGS uses queries as intent anchors to guide dynamic user interest generation based on comprehensive user historical data via the diffusion process.}
    \label{fig:modeling}
\end{figure}

Despite their success, we note that in prior DLRMs and GRMs, explicit modeling of the relationship between user input queries and historical behavior sequences remains underexplored.
For example, QIN~\cite{guo2023query} (as a representative DLRM) introduces a query-driven retrieval mechanism in their GSU module to identify ``relevant'' historical behaviors, and relies on external models to extract representations of user queries and target items, as illustrated in Figure~\ref{fig:modeling}(a).
On the other hand, GRMs~\cite{han2025mtgr,zhai2024actions} tend to treat queries as ordinary tokens that are concatenated with historical behavior sequences (see Figure~\ref{fig:modeling}(b)).
As a result, both DLRMs and GRMs struggle to accurately capture user instant intents, as they fail to explicitly align query semantics with the contextual nuances of historical behaviors.

Our insight to this problem is twofold. First, user input queries should serve as intent anchors, playing a dominating role in the subsequent interest extraction process, rather than interacting on equal footing (then buried) with the vast volume of user historical behaviors.
Second, current DLRMs and GRMs typically ignore ``irrelevant'' items, as these models either rely on attention mechanisms to integrate previously clicked items, assigning low attention weights to irrelevant ones, or directly filter them out using the GSU-ESU two-stage process.
However, such ``irrelevant'' items can provide valuable information to better characterize user profiles, e.g., reflecting user consumption levels.

In this paper, we aim to learn an effective representation of user interests by prioritizing user query information and mining useful item information from users' long-term, noisy behavioral data.
%
Motivated by the potential of diffusion models~\cite{ho2020denoising}, we propose a user interest modeling method (called {\bf DiffusionGS}) based on the controllable diffusion process (see Figure~\ref{fig:modeling}(c)).
Our model is implemented based on the GRM framework and consists of multiple stacked transformer layers~\cite{vaswani2017attention}.
We propose to leverage the user queries as conditional inputs to control the generation of user interests.
By performing noise addition and denoising processes in the behavioral sequences, we are able to extract more robust user interest representations, thereby enhancing the model’s capability to predict users' dynamic interests under current requests.
In addition, within the transformer architecture, we propose a \textbf{U}ser-aware \textbf{D}enoising \textbf{L}ayer (\textbf{UDL}) to integrate user-specific information into the denoising and interest extraction processes, thereby optimizing the distribution of attention weights and enhancing personalized ranking performance.

In summary, the contributions of this work are as follows.
\begin{itemize}
    \item We propose a novel \textbf{DiffusionGS} framework for capturing dynamic user interest shifts accurately. \textbf{DiffusionGS} combines the excellent scalability of generative models with the controllable generation capability of diffusion models, where user dynamic interest representations are generated based on comprehensive user historical behavior sequences, conditioned on the contextual information of user instant queries.
    \item We propose a \textbf{U}ser-aware \textbf{D}enoising \textbf{L}ayer (\textbf{UDL}), which enables the model to dynamically adjust attention weight distributions for personalized ranking by leveraging rich interactions between user profiles and historical behavior sequences.
    \item Extensive offline and online experiments demonstrate the effectiveness of \textbf{DiffusionGS}. We also validate the scaling law of the diffusion-based generative paradigm in search-ranking tasks.
\end{itemize}
\section{Related Work}

\subsection{User Historical Behavior Sequence Modeling}

User historical behavior sequence modeling serves as the cornerstone of Deep Learning based Ranking Models (DLRMs), enabling dynamic capture of evolving preferences and intent shifts. Early approaches like DIN~\cite{zhou2018deep} adopt target-aware attention to dynamically weight relevant historical behaviors. SASRec~\cite{kang2018self} replaces target attention mechanisms with a Transformer-inspired self-attention architecture. To scale to lifelong behavior sequences, SIM~\cite{pi2020search} introduces a two-stage retrieval paradigm, including a General Search Unit (GSU) rapidly filters top-K relevant behaviors, followed by an Exact Search Unit (ESU) that applies fine-grained target attention to model dynamic user interests. QIN~\cite{guo2023query} prioritizes query relevance in search scenarios by integrating a Query-Aware Relevance Unit to filter irrelevant items from behavioral sequences. To address the inconsistency of the GSU-ESU objective, TWIN~\cite{chang2023twin} designs a Consistency-Preserved GSU, which shares the ESU’s target-behavior relevance metrics. TWIN\_v2~\cite{si2024twin} employs hierarchical clustering with adaptive k-means to compress lifelong user behavior sequences into cluster-based representations, enabling scalable modeling of ultra-long sequences.

Target attention methods (e.g., DIN~\cite{zhou2018deep}) as well as approaches using queries or context to filter irrelevant items primarily focus on features of the most relevant items, thereby overlooking valuable information (e.g., long-term purchase value and style preferences) reflected by irrelevant ones, resulting in unsatisfactory search and ranking results.


\subsection{Generative Recommendation Models}
Recently, the migration of generative models to search and recommendation tasks has yielded advancements comparable to those observed in other AI fields. Generative models broadly fall into two categories: autoregressive models and diffusion models.

Autoregressive models generate data sequentially by predicting each next token conditioned on all preceding ones. One of the efforts in the recommendation task, HSTU~\cite{zhai2024actions}, uses a trillion‑parameter autoregressive architecture to model user-item interactions as sequences, achieving a promising performance gain. The follow-up MTGR~\cite{han2025mtgr} inherits the HSTU architecture and further preserves cross-feature richness from traditional DLRMs. TIGER~\cite{rajput2023recommender} formulates the recommendation task into semantic ID generation, by encoding items via RQ‑VAE~\cite{lee2022autoregressive} into discrete token sequences and predicting the next semantic ID autoregressively. HLLM~\cite{chen2024hllm} proposes a hierarchical dual‑LLM design, where an item‑level model is used to extract content features, and a user‑level model predicts next behavior given historical actions. 


Diffusion models learn a forward noising and reverse denoising process, starting from random noise and iteratively reconstructing structured data. A few methods have explored diffusion models for recommendation tasks. DiffuRec~\cite{li2023diffurec} corrupts only the target item embedding with Gaussian noise and then conducts reconstruction using a transformer-based approximator, achieving strong improvements across datasets. DiffRec~\cite{wang2023diffusion} adapts diffusion models for user-interaction generation, adding noise to sequences rather than pure item embeddings. DreamRec~\cite{yang2023generate} uses guided diffusion, where a transformer encodes the user history to steer the denoising process, generating an ``oracle'' item embedding that reflects true preference without relying on negative samples. 
Although these methods have demonstrated their effectiveness, we identify two key limitations. First, they are primarily designed for recommendation tasks and therefore underutilize the diffusion model’s potential for query-guided controllable generation. Second, they fall short in capturing the intrinsic user interests embedded within historical behavior sequences, thereby underleveraging the diffusion model’s strong capability to reconstruct complex data distributions.

\section{Preliminary}

\subsection{Problem Formulation}

In this section, we formally define the search task in the context of the Kuaishou E-commerce platform. Consider a general search engine involving the following sets:
\[
U = \{u \mid u \in \mathcal{U}\}, \quad
Q = \{q \mid q \in \mathcal{Q}\}, \quad
D = \{d \mid d \in \mathcal{D}\},
\]
where \(U\), \(Q\), and \(D\) denote the sets of users, queries, and documents, respectively. In the Kuaishou E-commerce scenario, the documents \(D\) correspond to candidate items available for purchase. Each item \(i \in D\) is associated with side information \(Side_i \in \mathbb{R}^{d_s}\), such as title, category, and price. For each user \(u \in U\), we observe a historical behavior sequence based on past interactions (clicks, purchases, or impressions). This behavior sequence is formally represented as:
\[
B_u = [(q^u_1, \{i^u_{1,1}, i^u_{1,2}, \dots\}),\; \dots,\; (q^u_{|B_u|}, \{i^u_{|B_u|,1}, \dots\})],
\]
where \(q^u_j\) denotes the \(j\)-th query issued by user \(u\), and \(\{i^u_{j,1}, i^u_{j,2}, \dots\}\) represents the sequence of items that were clicked, purchased, or shown in response to that query.

When a user enters a query in the search box and commits a request, the objective of our search system is to retrieve and rank items that are most likely to align with the user’s interests. Given a current query \(q\) issued by user \(u\), we define the \emph{target items} \(i_t\) as a candidate item pool to be ranked. In our E-commerce platoform, the main goal is to predict two key probabilities: (1) the click-through rate (CTR), i.e., the probability that the user will click on the target item, and (2) the conversion rate (CVR), i.e., the probability that the user will make a purchase after clicking on the target item.

\subsection{Comparison Between Search and Recommendation Tasks}

Although search and recommendation tasks both aim to present users with content of interest, they differ fundamentally in how user intent is expressed and utilized. The most critical distinction lies in the presence of an explicit query in search tasks. In recommendation scenarios, the system passively infers user preferences based on historical behavior, such as clicks or purchases, and proactively surfaces content that may be of interest. In contrast, search is an active user-driven process: users initiate interaction by issuing a query, which serves as a direct and explicit signal of intent.

This query plays a central role in guiding the retrieval and ranking of candidate items. It acts as a semantic filter that defines the user’s short-term goal, which differs from the long-term or general interests captured in the recommendation tasks. For example, consider a user who frequently browses electronic products and has a history of purchasing headphones. In the recommendation setting, the system might suggest new headphones or related accessories. However, if the same user enters the query \textit{``running shoes''} into the search box, the action clearly indicates a change in interest. A recommendation-style response (e.g., more electronics) would fail to satisfy the user's immediate need, highlighting the importance of incorporating the query context into search modeling.

While user behavior histories are important in both domains, the query constraint in search imposes an additional challenge: to effectively model not only the user's general preferences but also the intent expressed in the current query. This underscores the importance of query integration into personalized search models.

\section{Proposed Method}

\begin{figure*}
    \centering
    \includegraphics[width=\textwidth]{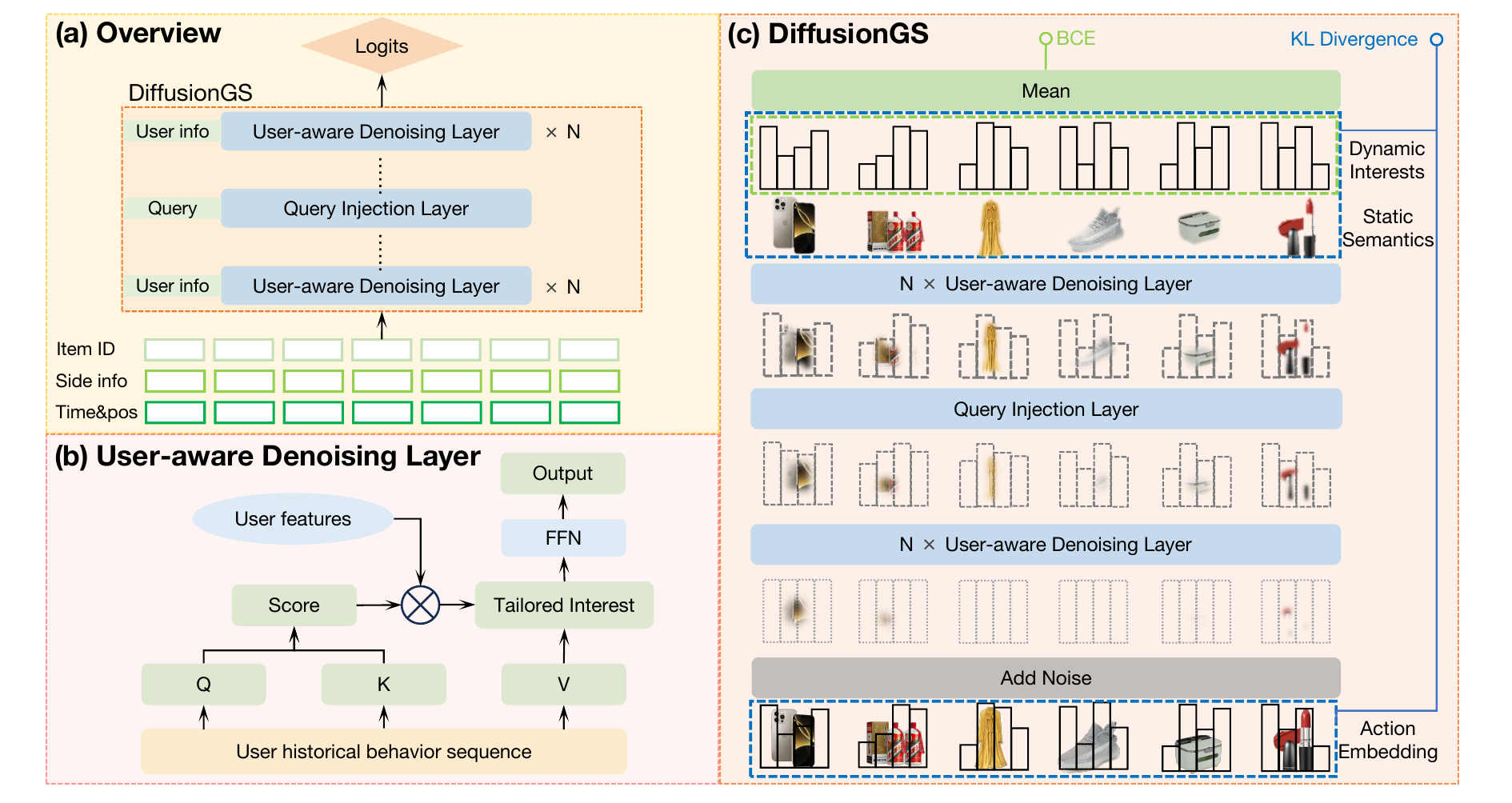}
    \caption{The proposed \textbf{DiffusionGS} framework. 
    (a) Overview of DiffusionGS: it takes item ID, side info, and temporal/positional embeddings as input and outputs logits for CTR and CVR prediction.
   (b) User-aware denoising layer: it integrates user features as gating signals to modulate attention weights for personalized ranking.
  (c) Denoising process:  we stack multiple user-aware denoising layers to progressively reconstruct dynamic user interests from behavior sequences, guided by a query injection layer conditioning on the current query.}
    \label{fig:framework}
\end{figure*}

We introduce \textbf{DiffusionGS}, a generative ranking framework that models the evolving interest distribution of each user by decoupling their historical interaction sequences under the guidance of current queries through a diffusion-based denoising process. Specifically, we conceptualize a user’s past behavior as the superposition of item-level semantic features (e.g., color, silhouette, price) and a dynamic user-specific interest distribution over these semantics (e.g., color bias, style affinity, willingness to pay). Importantly, disentangling these components is essential because interaction data often entangle latent factors, where some reflect genuine user intent, others arise from noise, such as popularity bias or transient trends. Isolated modeling of latent vectors enables more robust, interpretable, and generalizable preference representations~\cite{ma2019learning}. DiffusionGS achieves this disentanglement by denoising user historical behavior sequences to recover clean, dynamic interest representations, thereby improving future behavior prediction. 

Moreover, a user's dynamic interest should be conditioned on their current intent, as conveyed by the query. To that end, we guide the generation of user interest using query signals. Leveraging the controllable generation capabilities of diffusion models, we incorporate query information into the denoising process through conditional injection mechanisms.
This design enables explicit alignment between user interest modeling and the current query intent, thereby enhancing the relevance and specificity of the generated preference representations.

\subsection{Overview}

We illustrate the overall pipeline of our DiffusionGS model in Figure~\ref{fig:framework}. The input to DiffusionGS consists of a user's historical behavior sequence with both positive and negative user engagement, where each behavior at time step \(t\) is represented as \(I_t\), encoding both the item ID and its associated side information (e.g., title, category, and price).
To capture the temporal dynamics and ordering of user actions, which helps model evolving interests, recency effects, and sequential dependencies for a more accurate representation of user preference, we augment each behavior token with a timestamp embedding that encodes the absolute time of the action and a learnable positional embedding that captures the sequential order of actions within the historical behavior sequence.

Formally, the input token \(x_t\) at time step \(t\) is constructed as follows:
\begin{align}
I_t &= f\left(\mathrm{Concat}(\mathrm{ID}_{t},\; \mathrm{Side}_{t})\right), \\
x_t &= I_t + \mathrm{TimeEmb}(t) + \mathrm{PosEmb}(t),
\end{align}
where \(\mathrm{TimeEmb}(t)\) denotes the timestamp embedding, \(\mathrm{PosEmb}(t)\) is the learnable position embedding, and \(f(\cdot)\) is a transformation function (e.g., a multilayer perceptron) applied to the concatenation of the item ID and its side information \(\mathrm{Side}_{t} \in \mathbb{R}^{d_s}\). The detailed diffusion process and the injection mechanism of query condition into the model are described in Section~\ref{Diffusion Process} and Section~\ref{Condition Guided Generation}, respectively. After the denoising procedure in the diffusion process, our objective is to disentangle each item token \( x_t \) into two distinct latent representations: a static semantic component \( s_t \), capturing the intrinsic properties of the item, and a dynamic interest component \( d_t \), reflecting the user's evolving preferences. The average of the dynamic interest representations \( d_t \) is subsequently utilized to predict both CTR and CVR.

\subsection{Diffusion Process}
\label{Diffusion Process}

Motivated by the powerful capability of diffusion models to reconstruct structured data distributions from random Gaussian noise, we adopt them to decouple static semantic item features from dynamic user interests based on historical behavior sequences of users. 

Specifically, we employ a denoising diffusion probabilistic model (DDPM)~\cite{ho2020denoising} to model the complex generation process of dynamic user interests. The forward diffusion process gradually corrupts a clean input sequence by adding Gaussian noise over a series of time steps, eventually transforming it into a distribution similar to the Gaussian noise. Given an item token sequence \( \mathbf{x}_0 = \{x_1, x_2, \ldots, x_T\} \), the forward process produces a sequence of increasingly noisy versions \( \mathbf{x}_1, \mathbf{x}_2, \ldots, \mathbf{x}_T \) according to a pre-defined variance schedule \( \{\beta_t\}_{t=1}^T \), as
\begin{equation}
    q(\mathbf{x}_t \mid \mathbf{x}_{t-1}) = \mathcal{N}(\mathbf{x}_t; \sqrt{1 - \beta_t} \mathbf{x}_{t-1}, \beta_t \mathbf{I}),
\end{equation}
which leads to the marginal distribution:
\begin{equation}
    q(\mathbf{x}_t \mid \mathbf{x}_0) = \mathcal{N}(\mathbf{x}_t; \sqrt{\bar{\alpha}_t} \mathbf{x}_0, (1 - \bar{\alpha}_t) \mathbf{I}),
\end{equation}
where \( \alpha_t = 1 - \beta_t \) and \( \bar{\alpha}_t = \prod_{s=1}^t \alpha_s \).

Note that in our task setting, the primary objective is to estimate dynamic user interest distributions rather than to achieve high‑fidelity generation of item tokens. Consequently, we assert that moderate noise levels are sufficient to obscure spurious biases present in user behavior while still preserving essential semantic structure. As demonstrated in a previous study~\cite{li2023diffurec}, diffusion models with a moderate number of steps can effectively capture user preferences in sequential recommendation tasks, while significantly reducing computational overhead. We further evaluate the influence of different levels of noise intensity in Section~\ref{sec:intensity}.

The denoising process in our model is carried out in a stepwise manner using a neural network composed of stacked User-aware Denoising Layers (explained in Section~\ref{User-Aware self-attention module}). At each time step \( t \), the network predicts a clean representation \( \mathbf{x}_{t-1} \) from the noisy input \( \mathbf{x}_t \), guided by the query condition \( \mathbf{q} \). The complete generation of \( (\mathbf{s}_0, \mathbf{d}_0) \), including both static item semantics and dynamic user interests, is achieved by composing the intermediate transitions as:
\begin{align}
    P(\mathbf{x}_{t-1} \mid x_t, \mathbf{q}) &= P(\mathbf{s}_{t-1}+\mathbf{d}_{t-1} \mid x_t, \mathbf{q}) = f_t(\mathbf{x}_t, \mathbf{q}), \\
    P(\mathbf{s}_0, \mathbf{d}_0 \mid \mathbf{q}) &= f_0 \odot f_1 \odot \cdots \odot f_t(\mathbf{x}_t, \mathbf{q}),
\end{align}
where \( \odot \) denotes the element-wise transformation, and \( \mathbf{q} \) is the current query. Each function \( f_i \) corresponds to a denoising step that maps the noisy representation \( \mathbf{x}_i \) to a slightly cleaner form \( \mathbf{x}_{i-1} \), ultimately reconstructing the clean item token representation.
This iterative process is instantiated via DDPM, parameterized as:
\begin{equation}
    p_\theta(\mathbf{x}_{t-1} \mid \mathbf{x}_t, \mathbf{q}) = \mathcal{N}(\mathbf{x}_{t-1}; \boldsymbol{\mu}_\theta(\mathbf{x}_t, t, \mathbf{q}), \boldsymbol{\Sigma}_\theta(\mathbf{x}_t, t)),
\end{equation}
where \( \boldsymbol{\mu}_\theta \) and \( \boldsymbol{\Sigma}_\theta \) denote the predicted mean and covariance at time \( t \), respectively, and are learned functions conditioned on the noisy sample \( \mathbf{x}_t \), time step \( t \), and query condition \( \mathbf{q} \).

\subsection{Condition Guided Generation}
\label{Condition Guided Generation}

In controllable diffusion models, conditioning plays a pivotal role in guiding the generative process. There are three widely adopted strategies: (1) Additive Conditioning, (2) Concatenation Conditioning, and (3) Cross-attention Conditioning.

In our design, we adopt Additive Conditioning to incorporate query information \( \mathbf{q} \), in preference to concatenation-based or cross-attention approaches. Concretely, we inject the query encoding via element-wise addition between two user-aware denoising layers positioned in the middle of the network. This design allows the initial layers to focus on learning a rich feature representation of the user's historical behavior sequence, while the subsequent layers focus on generating query-guided dynamic user interest. Additive conditioning integrates the conditioning vector (e.g., a query embedding) into the same feature space as the original representation and merges it via element-wise addition. This operation directly modifies the latent feature map, thereby influencing all subsequent components of the model, including projection layers, normalization modules, and denoising dynamics.
Compared to concatenation conditioning (which simply increases input dimensionality without structurally altering the original features), and cross-attention conditioning (which modulates attention weights without fundamentally changing the underlying representations), additive conditioning offers a more deeply integrated and pervasive mechanism for injecting query-dependent signals into the model.
Refer to Section~\ref{Query Injection Comparison} for more details.

\subsection{User-aware Denoising Layer}
\label{User-Aware self-attention module}


In conventional self-attention layers, attention weights are computed solely based on the similarity between query and key vectors, and are uniformly applied to the value representations~\cite{vaswani2017attention}. Although this design has proven effective across a variety of tasks, it cannot incorporate external contextual signals such as user preferences or intent. As a result, it may fall short in personalized settings where different users require a differentiated focus on semantically distinct aspects of the same input. For example, in response to the query “sneakers”, a trend-conscious young user may prioritize features like “new arrivals” and “street style”, whereas a budget-sensitive middle-aged user may value attributes such as “discount” or “durability”. In such cases, a uniform attention mechanism cannot adequately account for these variations in user-specific relevance.

To address this limitation, we propose a User-aware Denoising Layer based on transformer layers~\cite{vaswani2017attention}, which introduces an auxiliary gating mechanism \(U\) to modulate the influence of each token in a user- and context-dependent manner. Meanwhile, we follow the design of HSTU~\cite{zhai2024actions} to apply a length-based scaling factor \(n\) to normalize attention scores, instead of using the conventional softmax normalization. The computation process can be written as:
\begin{align}
    Q &= XW_Q, \\ 
    K &= XW_K, \\ 
    V &= XW_V, \\
    H &= \left(\frac{QK^\top }{n} V \odot U \right),
\end{align}
where \(U \in \mathbb{R}^{n \times n}\) denotes a gating vector that encodes both user features (e.g., age and gender) and contextual information (e.g., timestamp and search scenario),
and \(n\) represents the length of the user's historical behavior sequence.

Unlike HSTU~\cite{zhai2024actions}, where \(U\) is parameterized and implicitly learned through transformation, our formulation derives \(U\) explicitly from user profiles and contextual signals, allowing the model to dynamically adjust its attention distribution in a personalized and interpretable manner.
The detailed composition of \(U\) is provided in the ``Gate features'' entry of Table~\ref{tab:side_info}.

\subsection{Objective Function and Training Loss}

We formulate our training objective as a combination of two loss components tailored to distinct but complementary goals: while the first component introduces a constraint to guide the diffusion-based generation process, ensuring that the synthesized representations retain fidelity to the original input token distribution, the second component aligns with standard practices in the prediction of CTR or CVR~\cite{zhou2018deep}, focusing on single-point prediction tasks. Given a specific target item, the model is trained to estimate the probability of user engagement (e.g., click or purchase).

\paragraph{Diffusion-Based Distribution Constraint.}
To ensure the fidelity of the generated sequence within the diffusion process, we introduce a Kullback–Leibler (KL) divergence loss. Specifically, we denote the reconstructed static tokens as $X^s$, the dynamically generated tokens as $X^d$, and the original input tokens as $X^{\text{input}}$. We constrain that the combined distribution of $X^s + X^d$ approximates the empirical distribution of $X^{\text{input}}$, which can be formulated as:
\begin{equation}
    \mathcal{L}_{\text{KL}} = D_{\text{KL}} \left( P(X^s + X^d) \parallel P(X^{\text{input}}) \right),
\end{equation}
where $P(\cdot)$ denotes the empirical distribution over the token embeddings. This KL divergence regularization enforces semantic consistency between the generated tokens and the original input, stabilizing the diffusion process and improving the quality and interpretability of the learned representations.
\paragraph{Pointwise Prediction Loss.}
The CTR (or CVR) task is treated as a binary classification problem for each single target item. Given the user context and the candidate target item, the model predicts a probability $\hat{y} \in [0, 1]$ that indicates the likelihood of a positive user interaction. The training signal is supervised using the binary cross-entropy (BCE) loss, defined as:
\begin{align}
    \hat{y} &= \sigma(\text{Average}(X^d)), \\
    \mathcal{L}_{\text{BCE}} &= - \left[ y \log(\hat{y}) + (1 - y) \log(1 - \hat{y}) \right],
\end{align} 
where $y \in \{0,1\}$ denotes the ground-truth label (i.e., 1 for click or purchase, 0 otherwise), and $\hat{y}$ is the model-predicted probability, computed from user dynamic interest $X^d$. $\sigma$ is the sigmoid activation function. This loss encourages the model to produce accurate point-wise predictions and aligns the learned embeddings with user preferences and item relevance.

\paragraph{Total Loss.}
The final training objective is a weighted sum of the diffusion regularization term and the pointwise prediction loss:
\begin{equation}
    \mathcal{L}_{\text{total}} = \mathcal{L}_{\text{BCE}} + \lambda \cdot \mathcal{L}_{\text{KL}},
\end{equation}
where $\lambda$ is a hyperparameter that balances predictive performance and generative quality.

\section{Experiments}

\subsection{Experiment Setup}
\subsubsection{Datasets}
For our generative search task, the ideal training and evaluation dataset should satisfy the following three criteria:

\paragraph{Inclusion of Search Queries:} The dataset should explicitly contain real user search queries. While the prior work~\cite{guo2023query} adopts recommendation datasets~\cite{mcauley2015image, harper2015movielens} and synthesizes queries using existing methods (e.g., as described in~\cite{van2016learning}), synthetic queries derived from item descriptions often lack the diversity, ambiguity, and intent-specific nuances of real user queries. This discrepancy can result in a discrepancy between training and real-world deployment scenarios~\cite{rahmani2025towards}. Hence, we prioritize datasets that include authentic search queries.

\paragraph{Rich User Behavioral History:} To accurately model users' evolving interests, the dataset should contain diverse and sufficiently long sequences of user historical behavior. However, most public search datasets~\cite{Sun2023KuaiSAR, mcauley2015image} fail to provide such information. For example, in the KuaiSAR dataset~\cite{Sun2023KuaiSAR}, the average length of a user’s behavioral history is only about 24, which is far from sufficient to reflect the complexity of real user preferences.

\paragraph{Availability of Rich Side Information:} Side information (e.g., item attributes, user profiles, and contextual metadata) is essential for learning robust and fine-grained item representations to enhance the model's generalization and personalized ranking performance. However, existing datasets such as the Taobao Search Dataset~\cite{hu2018reinforcement} only contain four item attributes beyond the query itself, posing significant challenges for effective personalization.

While existing publicly available datasets fall short of our requirements, we construct a new dataset collected from the Kuaishou E-commerce platform. Specifically, we sample user interaction data spanning 60 consecutive days as our training set and designate the data of the following day as the evaluation set. 
On average, the dataset contains approximately 14 million user interaction records per day, with each record comprising around 177 user click interactions.
This dataset offers a large-scale (e.g., 20 times larger than the KuaiSAR search dataset~\cite{Sun2023KuaiSAR}), real-world foundation for training and evaluating our generative search model. Example side information used in our dataset is listed in Table~\ref{tab:side_info}.

\begin{table}[htbp]
\centering
\caption{Side information examples in our dataset.}
\begin{tabular}{|p{2.5cm}|p{5.5cm}|}
\hline
\textbf{Features} & \textbf{Side Information} \\
\hline
User features & user ID, gender, age, location, ... \\
\hline
Item features & item ID, price, detail page view time, category, seller ID, ... \\
\hline
Query features & keyword, predicted age, predicted gender, ... \\
\hline
Context features & timestamp, search scene, client, ... \\
\hline
Gate features & user features, context features \\
\hline
\end{tabular}
\label{tab:side_info}
\end{table}

\subsubsection{Metrics}

In search and recommendation scenarios, the performance of ranking models is typically evaluated using metrics that assess the quality of the predicted ranking of target items concerning user interactions. Among these, we use the Area Under the ROC Curve (AUC) and Grouped AUC (GAUC), which are two widely adopted criteria to measure model effectiveness~\cite{ferri2011coherent}.

The AUC metric evaluates the probability that a randomly chosen positive instance (e.g., a clicked item) is ranked higher than a randomly chosen negative instance (e.g., a non-clicked item). Formally, for a set of predictions, AUC is defined as:
\begin{equation}
\mathrm{AUC} = \frac{1}{|P||N|} \sum_{p \in P} \sum_{n \in N} \mathbb{I}(s_p > s_n),
\end{equation}
where $P$ and $N$ denote the sets of positive and negative instances, respectively, $s_p$ and $s_n$ are the predicted scores for a positive instance $p$ and a negative instance $n$, and $\mathbb{I}(\cdot)$ is the indicator function.

However, AUC computed on the entire dataset may overlook request-level performance disparities. To address this, GAUC is introduced to capture the average AUC across user requests, which is defined as:
\begin{equation}
\mathrm{GAUC} = \frac{1}{N_R}\sum_{r \in R} \mathrm{AUC}_r,
\end{equation}
where $R$ is the set of requests, $N_R$ is the number of requests, and $\mathrm{AUC}_r$ is the AUC computed via interactions in request $r$. GAUC provides a more accurate reflection of model performance in personalized ranking settings, as it considers inter-request variance.

\subsubsection{Parameter Setting}

We train our model using the Adam~\cite{adam2014method} optimizer. The training task is conducted on four NVIDIA A100 GPUs, with each GPU handling a batch size of 4096. The learning rate is set to 0.00025 during the training process. The diffusion process is carried out in 50 steps for both noise addition and denoising. A linear sampling schedule is used, with a noise intensity ranging from 0.005 to 0.01.

\subsubsection{Baselines}

We compare our DiffusionGS model with three DLRMs, including SASRec~\cite{kang2018self}, SIM~\cite{pi2020search}, and QIN~\cite{guo2023query}, and two GRMs: HSTU \cite{zhai2024actions} and MTGR \cite{han2025mtgr}.

\textbf{SASRec}~\cite{kang2018self} adapts the Transformer self‐attention mechanism for sequential recommendation, effectively capturing the long‐term dependencies in user behavior.

\textbf{SIM}~\cite{pi2020search} handles ultra-long sequences through a two-stage retrieval-attention framework (GSU for coarse search, ESU for precise modeling).

\textbf{QIN}~\cite{guo2023query} enhances intent-aware item ranking by integrating queries with user behavior history via a relevance search unit, filtering user interaction sequences by query relevance.

\textbf{HSTU}~\cite{zhai2024actions} employs an autoregressive architecture to model user–item interactions as sequential token generation, treating recommendation as a sequence-to-sequence transduction task.

\textbf{MTGR}~\cite{han2025mtgr} extends the HSTU architecture by reintroducing traditional DLRM-style cross features and a Group-Layer Normalization mechanism.

\subsection{Results}
\begin{table}[htbp]
\centering
\vspace{-5pt}
\caption{Comparison with state-of-the-art DLRM and GRM methods on our large-scale real-world dataset.}
\label{tab: overall performance}
\begin{tabular}{lcccccc}
\toprule
\multirow{2}{*}{Methods} & \multicolumn{2}{c}{CTR} & \multicolumn{2}{c}{CVR} \\
\cmidrule(lr){2-3} \cmidrule(lr){4-5}
 & AUC$\uparrow$ & GAUC$\uparrow$ & AUC$\uparrow$ & GAUC$\uparrow$ \\
\midrule
SASRec~\cite{kang2018self}    & 0.7320 & 0.6301 & 0.8547 & 0.5886 \\
SIM~\cite{pi2020search}      & 0.7301 & 0.6290 & 0.8545 & 0.5901 \\
QIN~\cite{guo2023query}       & 0.7313 & 0.6303 & 0.8544 & 0.5890 \\
HSTU~\cite{zhai2024actions}      & 0.7268 & 0.6236 & 0.8493 & 0.5812 \\
MTGR~\cite{han2025mtgr}      & 0.7250 & 0.6223 & 0.8491 & 0.5823 \\
\textbf{DiffusionGS}  & \textbf{0.7502} & \textbf{0.6403} & \textbf{0.8635} & \textbf{0.5963} \\
\bottomrule
\end{tabular}
\end{table}

We first compare our proposed DiffusionGS with five representative baseline methods on both the CTR and the CVR prediction tasks. As shown in Table~\ref{tab: overall performance}, DiffusionGS consistently achieves the best performance on all metrics. Specifically, on the CTR prediction task, DiffusionGS obtains the highest AUC (0.7502) and GAUC (0.6403), outperforming all baselines by a significant margin. Similarly, on the CVR prediction task, DiffusionGS achieves the highest AUC (0.8635) and GAUC (0.5963), demonstrating its ability to generalize well across different tasks. These results suggest that DiffusionGS can effectively decouple the dynamic user interest from the user's historical behavior sequence based on underlying user intention, thereby enhancing prediction accuracy.

Among the baseline models, DLRM-based methods (i.e., SASRec~\cite{kang2018self}, SIM~\cite{pi2020search}, and QIN~\cite{guo2023query}) demonstrate promising performance across the reported metrics. SASRec achieves the second-best AUC results in both the CTR and CVR prediction tasks, while QIN and SIM produce the second-best GAUC scores in the CTR and CVR tasks, respectively.
These results indicate that models designed to capture user behavior sequences are generally effective in modeling user preferences and behavior patterns. However, they may still struggle to capture fine-grained conditional dependencies (e.g., query context), which results in their inferior performance compared to our proposed method.
On the other hand, previous GRM-based methods (i.e., HSTU~\cite{zhai2024actions} and MTGR~\cite{han2025mtgr} perform relatively worse.
These results suggest that their designs for autoregressive tasks in a sequential modeling framework may not be optimal for downstream deep learning-based search tasks.


\subsection{Internal Analysis}

\subsubsection{Query Conditioned Generation}

\begin{figure}
    \centering
    \includegraphics[width=\linewidth]{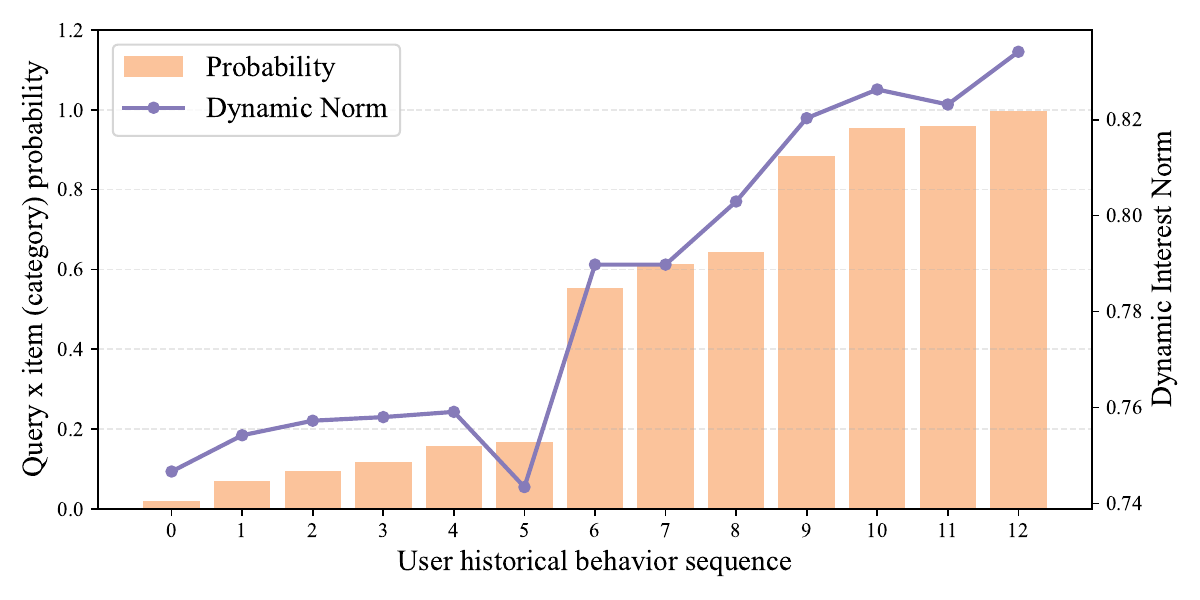}
    \caption{
    Visualization of the alignment between captured dynamic user interest and user queries. The bar plot shows predicted click probabilities by item category, and the line plot depicts the $\ell_2$ norm of the dynamic interest vector for each item. The user’s behavior sequence is sorted by ascending click probability.}
    \label{fig:vis}
\end{figure}

We first validate that the learned dynamic user interest is indeed guided by query signals, by visualizing the relationship between dynamic user interest and the query in Figure~\ref{fig:vis}. We use the $\ell_2$ norm of the dynamic user interest vector as a proxy for the user’s attention to a particular item, where a higher norm indicates stronger interest. For the query, we evaluate its influence by computing the predicted probability that a user will click on an item from each category. Specifically, when a user clicks on an item, we identify the corresponding item category and use the prediction model to estimate the click probability conditioned on the query. In Figure~\ref{fig:vis}, the bar plot represents the predicted query-item (category) click probabilities, while the line plot shows the $\ell_2$ norm of the dynamic interest vectors for each item. 

A general trend observed in Figure~\ref{fig:vis} is that categories more closely aligned with the query, reflected by higher predicted click probabilities, tend to have larger dynamic interest norms, indicating greater user interest. Beyond this global trend, another noteworthy observation is that DiffusionGS does not completely disregard items that are less relevant to the query. Instead, it incorporates them into the final representation, supporting our insight that seemingly irrelevant items can still provide valuable signals for predicting a user's next action. Furthermore, even among items with similar predicted category-level relevance, DiffusionGS is able to capture subtle distinctions in user preference, as evidenced by the varying magnitudes of their dynamic interest norms.

\subsubsection{Query Injection Comparison}
\label{Query Injection Comparison}

We then examine the impact of different query injection methods on prediction performance. Commonly used conditioning strategies in the image generation domain include additive conditioning, concatenation-based conditioning, and cross-attention-based conditioning. 

\emph{Additive Conditioning} introduces conditional signals by injecting them into intermediate feature maps via element-wise addition or zero-initialized residual paths, as exemplified by ControlNet~\cite{zhang2023adding} and T2I-Adapter~\cite{mou2024t2i}. This approach enables seamless integration of condition priors without altering the input dimensionality.

\emph{Concatenation Conditioning} expands the input space by appending condition representations—such as low-resolution images or structural maps—either along the channel or spatial dimension. For instance, SR3~\cite{saharia2022image} concatenates an upsampled low-resolution image with the noisy input to guide super-resolution, while HumanSD~\cite{ju2023humansd} integrates pose heatmaps to steer human image generation.

\emph{Cross-attention Conditioning} incorporates external signals via attention modules, where conditional inputs are treated as queries or keys within transformer-style attention blocks. Representative examples include IP-Adapter~\cite{ye2023ip} and GLIGEN~\cite{li2023gligen}, which align text or image conditions with the generative process through attention-based fusion.

We find that additive conditioning has a more direct influence on all downstream components of the model, leading to a more pronounced effect on overall performance. Consequently, we adopt additive conditioning as the default approach in our method, while also evaluating the other two conditioning methods for comparison.

The results of the CVR prediction task are presented in Table~\ref{tab: Query injection comparison},
where the ``DiffusionGS-cat'' model refers to the concatenation-based conditioning variant, in which queries are concatenated with the user behavior sequence before being input into the network. The ``DiffusionGS-cross'' model denotes the cross-attention-based conditioning variant, in which the final self-attention layer is replaced by a cross-attention layer with the user queries used as the query (Q) input, and the user behavior sequence retained as the key (K) and value (V) inputs. 

As shown in Table~\ref{tab: Query injection comparison}, our DiffusionGS model with additive conditioning outperforms the other two variants, achieving an AUC of 0.8635 and GAUC of 0.5963. In contrast, DiffusionGS-cat and DiffusionGS-cross yield slightly lower performance, with AUC/GAUC scores of 0.8615/0.5935 and 0.8614/0.5936, respectively. These results verify our hypothesis that additive conditioning provides more effective integration of query signals, leading to improved predictive performance.

\begin{table}[htbp]
  \centering
  \vspace{-5pt}
  \caption{Comparison of condition injection variants on the CVR task.}
  \begin{tabular}{lcc}
    \toprule
    Methods & AUC$\uparrow$ & GAUC$\uparrow$ \\
    \midrule
    DiffusionGS-cat & 0.8615 & 0.5935 \\
    DiffusionGS-cross & 0.8614 & 0.5936 \\
    \textbf{DiffusionGS} & \textbf{0.8635} & \textbf{0.5963} \\
    \bottomrule
  \end{tabular}
  \label{tab: Query injection comparison}
\end{table}

\subsubsection{Impact of User-Aware Denoising Layer}

\begin{table}[htbp]
  \centering
  \vspace{-5pt}
  \caption{Impact of the User-aware denoising layer, demonstrated in the CVR task.}
  \begin{tabular}{lcc}
    \toprule
    Methods & AUC$\uparrow$ & GAUC$\uparrow$ \\
    \midrule
    DiffusionGS-learntU & 0.8615 & 0.5933 \\
    DiffusionGS-NoU & 0.8616 & 0.5928 \\
    \textbf{DiffusionGS} & \textbf{0.8635} & \textbf{0.5963} \\
    \bottomrule
  \end{tabular}
  \label{tab: Impact of USAU}
\end{table}

We now evaluate the effectiveness of the proposed User-aware Denoising Layer (UDL) with the CVR task. We compare the full DiffusionGS model (which applies user profiles as gating information) with two variants (i.e., ``DiffusionGS-learntU'' and ``DiffusionGS-NoU'' in Table~\ref{tab: Impact of USAU}). The ``DiffusionGS-learntU'' variant is inspired by HSTU~\cite{zhai2024actions}, where gating weights are implicitly learned from the historical behaviors of users. In contrast, ``DiffusionGS-NoU'' removes the gating mechanism directly and relies solely on a standard QKV self-attention mechanism. The results are presented in Table~\ref{tab: Impact of USAU}. Both ``DiffusionGS-learntU'' and ``DiffusionGS-NoU'' achieve less satisfactory AUC scores of approximately 0.8615, while our DiffusionGS model achieves an AUC of 0.8635.
For the GAUC metric, although DiffusionGS-learntU slightly outperforms DiffusionGS-NoU, it still underperforms compared to our DiffusionGS model.
These results highlight the effectiveness of the User-aware Denoising Layer in leveraging user-specific characteristics to enhance personalized ranking performance.

\subsubsection{Noise Intensity Comparison}
\label{sec:intensity}

\begin{table}[htbp]
  \centering
  \vspace{-5pt}
  \caption{Comparison of the noise intensity on the CVR task.}
  \begin{tabular}{lcc}
    \toprule
    Methods & AUC$\uparrow$ & GAUC$\uparrow$ \\
    \midrule
    DiffusionGS-Nonoise & 0.8616 & 0.5943 \\
    DiffusionGS-Noise2000 & 0.8617 & 0.5938 \\
    \textbf{DiffusionGS} & \textbf{0.8635} & \textbf{0.5963} \\
    \bottomrule
  \end{tabular}
  \label{tab: Noise intensity comparison}
\end{table}

Recall that we adopt 50 steps for both noise injection and denoising (in Section~\ref{Diffusion Process}), as we hypothesize that moderate noise levels are sufficient to obscure spurious biases in user behavior while preserving the essential semantic structure. 
To validate the effectiveness of this design choice, we conduct an ablation study with two variants: ``DiffusionGS-Nonoise'' and ``DiffusionGS-Noise2000''. The ``DiffusionGS-Nonoise'' variant removes the diffusion process, eliminating both noise addition and removal. In contrast, the ``DiffusionGS-Noise2000'' variant increases the number of diffusion steps to 2000 and increases the maximum noise intensity to 0.1,
which results in a noise-perturbed vector that effectively obscures the user’s historical behavior.

Table~\ref{tab: Noise intensity comparison} reports the results of the noise intensity comparison. We observe that both the absence of noise (i.e., ``DiffusionGS-Nonoise'') and the excessive use of noise (i.e., ``DiffusionGS-Noise2000'') lead to suboptimal performance, compared to using a moderate level of noise in our model. These results verify our hypothesis that neither clean nor heavily damaged user historical behavior data is suitable for reconstructing dynamic user interests.

\subsubsection{Scaling law of Diffusion-GS}

\begin{figure}
    \centering
    \includegraphics[width=\linewidth]{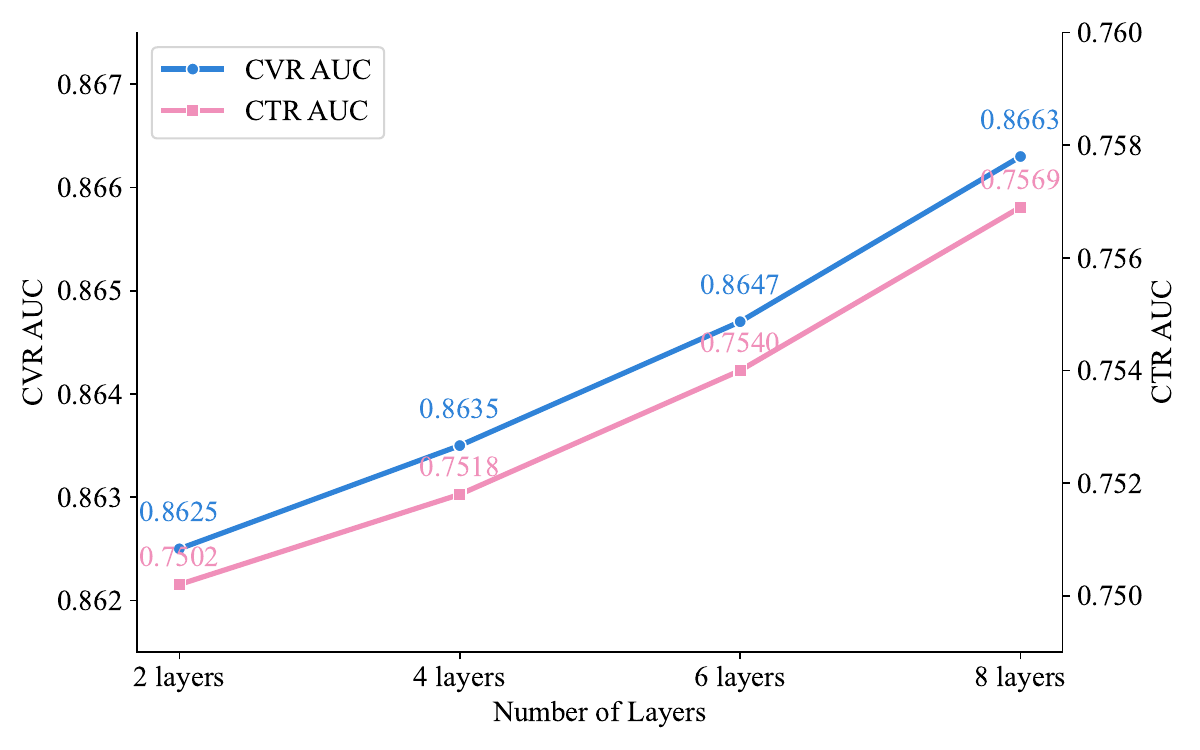}
    \caption{Scaling law of Diffusion-GS. The model's performance improves as the number of parameters increase.}
    \label{fig:scale}
\end{figure}

Scaling laws describe the empirical relationship between model performance and the scale of model parameters. In recent years, many studies~\cite{han2025mtgr, zhai2024actions, yan2025unlocking} in the recommendation domain have validated the presence of scaling laws, demonstrating that appropriately increasing the capacity of ranking models can lead to consistent performance improvements.

Motivated by these findings, we investigate whether the scaling law holds for our model by varying the depth of the UDL layers. Specifically, we increase the number of UDL layers from 2 to 8 and evaluate the model on the CVR task. The results, as shown in Figure~\ref{fig:scale}, indicate a clear upward trend: the performance improves from 0.8625 to 0.8663, yielding a 0.44\% absolute gain. This empirical evidence supports that scaling up the UDL layers contributes positively to model effectiveness, validating the applicability of the scaling law in our diffusion-based framework.

\subsection{Online Model Deployment}

\begin{figure}
    \centering
    \includegraphics[width=0.75\linewidth]{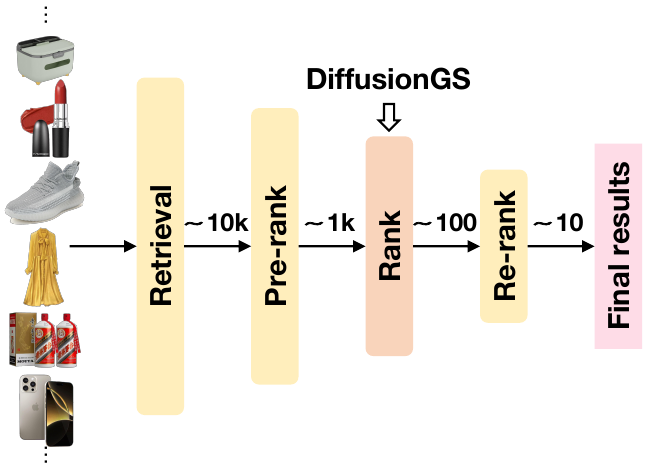}
    \caption{Illustration for the cascade framework in online search and recommendation systems. Our DiffusionGS model works on the ranking stage.}
    \label{fig:cascade}
\end{figure}

Figure~\ref{fig:cascade} illustrates how cascade models in search and recommendation balance efficiency and effectiveness through sequential stages: a retrieval stage first narrows candidates to $\sim$10,000 items; the pre-ranking stage further reduces the number of candidates to $\sim$1,000 using moderately complex models; ranking is then applied through sophisticated models to select the top hundreds; and finally the post-ranking stage determines ordering with business constraints.

Our DiffusionGS model is currently deployed on the ranking stage. In offline experiments, we adopt 50 steps for both the noise addition and denoising phases. For the online deployment, where system latency plays a critical role in shaping user experience, we reduce the computational overhead by performing only a single denoising step to generate the final output. Empirically, we observe that this simplification leads to negligible performance degradation (e.g., about 0.01\% decrease in AUC) compared to the full 50-step denoising process, while significantly improving inference efficiency.
This verifies that single-step denoising is an effective design choice in the context of online search and recommendation, as it does not require the generation of high-frequency, lossless texture details as in image generation tasks.

\subsection{Online A/B Test}

We have deployed our DiffusionGS model in the Kuaishou E-Commerce Search Platform, to evaluate the online effectiveness against our online base model (which is a traditional DLRM similar to SIM~\cite{pi2020search}). We have conducted an online experiment for 14 days, involving billions of user requests. We have observed obvious improvements in four main metrics, as summarized in Table~\ref{tab: online}.
The \#Total Orders and \#GMV (Gross Merchandise Value) metrics measure the user engagement on our online shopping platform, while online CTR and CVR measure the effectiveness of the models. This online experiment demonstrates the superior performance of DiffusionGS in real-world applications, providing a more engaging shopping experience for billions of users.

\begin{table}[htbp]
  \centering
  \caption{Online A/B test of DiffusionGS.}
\resizebox{0.95\linewidth}{!}{
  \begin{tabular}{lcccc}
    \toprule
    Methods & \#Total GMV$\uparrow$ & \#Total Orders$\uparrow$ & CTR$\uparrow$ & CVR$\uparrow$ \\
    \midrule
    Base (DLRM) & - & - & - & -\\
    DiffusionGS & \textbf{+1.550\%} & \textbf{+1.339\%} & \textbf{+0.732\%}  & \textbf{+0.855\%} \\
    \bottomrule
  \end{tabular}}
  \label{tab: online}
\end{table}

\section{Conclusion}

In this paper, we have proposed DiffusionGS, a novel generative ranking model. Based on conditional injection techniques, our model optimizes the user interest representation with the utilization of user queries.
By introducing the diffusion process into our task, DiffusionGS can extract accurate user interest from noisy historical behaviors. 
In addition, we design a user-aware denoising layer (UDL) to integrate user-specific profiles into search and ranking. Built upon UDLs, our model enjoys better capacity for personalized ranking.
Compared to state-of-the-art DLRMs and GRMs, DiffusionGS achieves notable improvements in extensive offline experiments. Furthermore, the effectiveness of our model is validated through online A/B testing, demonstrating its practical value in real-world applications.

Despite its promising performance, DiffusionGS, tailored for ranking tasks, still relies on the conventional cascade pipeline to narrow down the scope of candidate items. This non-end-to-end optimization paradigm may limit the improvement of user satisfaction and overall system effectiveness. In our future work, we envision extending DiffusionGS toward a generative paradigm that directly models the user-to-item generation process. 
This could potentially eliminate the need for multistage pipelines and lead to a more coherent and adaptive searching/recommendation framework. We hope that the insights from DiffusionGS may inspire new paradigms for personalized modeling in future search and recommendation domains.

\bibliographystyle{ACM-Reference-Format}
\bibliography{reference}
\end{document}